\documentclass[aoas,MSNbibl,nameyear,dvips]{arximspdf}
\usepackage{graphicx}
\doi{10.1214/14-AOAS718} 
\volume{8}
\issue{2}
\pubyear{2014}
\firstpage{1095}
\lastpage{1118}

\makeatletter
\def\geqslant{\geq}
\def\leqslant{\leq}

\newcommand{\eqref}[1]{(\ref{#1})}
\newcommand{\disp}{ }
\newcommand{\dispf}{ \frac}
\makeatother

\begin{document}
\begin{frontmatter}

\title{Analysis of multiple sclerosis lesions via spatially varying coefficients}
\runtitle{Analysis of MS lesions}

\begin{aug}
\author[a]{\fnms{Tian}~\snm{Ge}\thanksref{m1,m2}\ead[label=e1]{tge@fudan.edu.cn}},
\author[b]{\fnms{Nicole}~\snm{M\"{u}ller-Lenke}\thanksref{m3}\ead[label=e2]{nmueller@uhbs.ch}},
\author[b]{\fnms{Kerstin}~\snm{Bendfeldt}\thanksref{m3}\ead[label=e3]{kerstin.bendfeldt@unibas.ch}},
\author[c]{\fnms{Thomas E.}~\snm{Nichols}\thanksref{m2,t1}\ead[label=e4]{t.e.nichols@warwick.ac.uk}}
\and
\author[d]{\fnms{Timothy D.}~\snm{Johnson}\corref{}\thanksref{m4,t1}\ead[label=e5]{tdjtdj@umich.edu}}
\runauthor{T. Ge et al.}

\thankstext{t1}{Supported in part by NIH Grant
5-R01-NS-075066-02 (TDJ, TEN), the United Kingdom's
Medical Research Council Grant G0900908 (TEN) and the Wellcome Trust (TEN).
The work presented in this manuscript represents the views of the
authors
and not necessarily that of the NIH, UKMRC or the Wellcome Trust.}

\affiliation{Fudan University\thanksmark{m1}, University of
Warwick\thanksmark{m2},
University Hospital Basel\thanksmark{m3}\break  and University of
Michigan\thanksmark{m4}}

\address[a]{T. Ge\\
Centre for Computational Systems Biology\\
School of Mathematical Sciences\\
Fudan University\\
Shanghai 200433\\
China\\
and\\
Department of Computer Science\\
University of Warwick\\
Coventry CV4 7AL\\
United Kingdom\\
\printead{e1}}

\address[b]{N. M\"{u}ller-Lenke\\
K. Bendfeldt\\
Medical Image Analysis Center (MIAC)\\
University Hospital Basel\\
CH-4031 Basel\\
Switzerland\\
\printead{e2}\\
\phantom{E-mail:\ }\printead*{e3}}

\address[c]{T.~E. Nichols\\
Department of Statistics and\\
\quad Warwick Manufacturing Group\\
University of Warwick\\
Coventry CV4 7AL\\
United Kingdom\\
\printead{e4}}

\address[d]{T.~D. Johnson\\
Department of Biostatistics\hspace*{48pt}\\
School of Public Health\\
University of Michigan\\
1415 Washington Heights\\
Ann Arbor, Michigan 48109\\
USA\\
\printead{e5}}
\end{aug}

\received{\smonth{1} \syear{2013}}
\revised{\smonth{7} \syear{2013}}

%
\begin{abstract}
Magnetic resonance imaging (MRI) plays a vital role
in the scientific investigation and clinical management of multiple sclerosis.
Analyses of binary multiple sclerosis lesion maps are typically ``mass
univariate'' and conducted with standard linear models
that are ill suited to the binary nature of the data and ignore the
spatial dependence between nearby voxels (volume elements).
Smoothing the lesion maps does not entirely eliminate the non-Gaussian
nature of the data
and requires an arbitrary choice of the smoothing parameter.
Here we present a Bayesian spatial model to accurately model binary
lesion maps
and to determine if there is spatial dependence between lesion location
and subject specific covariates
such as MS subtype, age, gender, disease duration and disease severity measures.
We apply our model to binary lesion maps derived from $T_{2}$-weighted
MRI images
from 250 multiple sclerosis patients classified into five clinical subtypes,
and demonstrate unique modeling and predictive capabilities over
existing methods.
\end{abstract}

%

\end{frontmatter}

\section{Introduction}\label{sec1}

Multiple sclerosis (MS) is an autoimmune disease of the central nervous system
characterized by neuronal demyelination that results in brain and
spinal cord lesions.
These lesions can appear throughout the brain but are more prevalent in
white matter.
Damage to the myelin and axons, the ``wires'' of the central nervous system,
affects the ability of nerve cells to communicate effectively and leads
to deficits
in the motor, sensory, visual and autonomic systems [\citet{Compston2008}].
Clinical symptoms of MS occur in episodic acute periods of attacks
(relapsing forms),
in gradual progressive deterioration of neurologic function
(progressive forms) or in a combination of both [\citet{Lublin1996}].
Patients are categorized into different MS subtypes based on these
clinical disease courses.
However, the progression of the disease and the formation of lesions
are highly heterogeneous both within and between individuals.

MRI is an established tool in the diagnosis of MS and in monitoring its
evolution [\citet{Bakshi2008,Filippi2011}].
A single MRI scanner can produce a range of different types of images.
$T_1$-weighted MRI images show white matter as most intense, gray
matter darker and cerebral spinal fluid darkest.
$T_2$-weighted MRI images show cerebral spinal fluid as most intense,
gray matter darker and white matter darkest;
air has no signal in either type of image.
In MS, $T_{1}$-weighted images identify areas of permanent axonal
damage that appear as hypointense ``black holes.''
$T_{2}$-weighted images show both new and old lesions as hyperintense regions.
These MRI scans provide complementary information about the nature of
MS and
are important tools used to monitor disease course in both time and
space [\citet{Neema2007}].
$T_{1}$ and $T_{2}$ images also support various approaches in lesion
detection, lesion segmentation [\citet{Anbeek2004}],
patient phenotyping and patient classification [\citet{Bakshi2008}].
For quantitative analysis of MS lesions from MRI scans, researchers
create lesion maps, binary images that mark the exact location of the lesions.
After registering all subjects to a common anatomical atlas,
they create lesion probability maps (LPM) that show the empirical
lesion rate at each voxel (or volume element).

Despite the importance of MRI for management of MS, clinically observed
disease progression correlates only poorly with conventional MRI findings;
this is so notable that some researchers call the lack of such
associations a paradox [\citet{Kacar2011}].
Possible reasons for this paradox include an underestimation of brain
damage by conventional MRI
and a lack of histopathological specificity of MRI findings [\citet
{Barkhof2002}].
This, however, has prompted structural investigations of the so-called
normal-appearing brain tissue outside the MR visible lesions [\citet
{Vrenken2006}].
For example, another type of MRI, diffusion tensor imaging (DTI),
is used for measuring changes in the normal appearing brain tissues
[\citet{Rovaris2005}]
and is used to investigate the relationship between diffusion abnormalities
and clinical disabilities in MS patients [\citet
{Werring1999,Filippi2001,Roosendaal2009,Goldsmith2011}].

To better detect and understand MRI-clinical associations, a number of
authors have focused on voxel-by-voxel analyses of LPMs.
Some studies compare distribution of patterns of lesions from different
MS subtypes [\citet{Holland2012,Filli2012}].
Others have proposed correlating lesion map data with different types
of clinical deficits;
``Voxel-based lesion-symptom mapping'' (VLSM) [\citet{Bates2003}] is the name
given to voxel-by-voxel modeling of tissue damage with behavioral and
clinical correlates.
However, these methods are all ``mass univariate'' and ignore the
spatial dependence between nearby voxels.
These methods also use standard linear models that are inappropriate
for binary data.
Often, researchers smooth the lesion maps, but this does not completely
eliminate the non-Gaussian nature of the data
and requires an arbitrary choice of the smoothing parameter
[\citeauthor{Charil2003} (\citeyear{Charil2003,Charil2007})
\citet{Kincses2011,Dalton2012}].
For example, \citet{Charil2003} perform voxel-wise linear regressions
between lesion probability and different clinical disability scores
to identify the regions preferentially responsible for different types
of clinical deficits.
\citet{Charil2007} correlate focal cortical thickness
with white matter lesion load and with MS disability scores.
In both studies, the authors apply an arbitrary smoothing kernel and
perform the analyses independently at each voxel or cortical vertex.

Our motivation for this work is twofold:
(1) to appropriately model these binary lesion maps and model the local
spatial dependence and, more importantly,
(2) to obtain sensitive inferences on the presence of spatial
associations between lesion location and subject specific covariates
such as MS subtype, age, gender, disease duration (DD) and
disease severity as measured by the Expanded Disability Status Scale
(EDSS) score and the Paced Auditory Serial Addition Test (PASAT) score.
To this end, we propose a Bayesian spatial model of lesion maps.
In particular, we propose a spatial generalized linear mixed model with
spatially varying coefficients.
The spatially varying coefficients are latent spatial processes (or fields).
We model these processes jointly using a multivariate pairwise
difference prior model,
a particular instance of the multivariate conditional autoregressive
model [\citeauthor{Besag1974} (\citeyear{Besag1974,Besag1993}),
\citet{Mardia1988}].
Our model fully respects the binary nature of the data and the spatial
structure of the lesion maps
as opposed to the aforementioned mass univariate methods.
Furthermore, our model produces regularized (smoothed) estimates of
lesion incidence without an arbitrary smoothing parameter.
Our model also allows for explicit modeling of the spatially varying
effects of covariates
such as age, gender and disabilities scores (e.g., the EDSS and PASAT scores),
producing spatial maps of these effects and their significance,
as well as the (scalar) effect of spatially varying covariates such as
the fraction of white matter in each voxel.

The idea of spatial modeling with spatially varying coefficient
processes traces back to \citet{Gelfand2003}.
They model the coefficient surface as a realization of a Gaussian
spatial process.
The correlation function of the Gaussian process determines the
smoothness of the process.
However, difficulties arise when the number of sites is large.
In particular, inversion of the correlation matrix is computationally
infeasible.
To overcome this difficulty, \citet{Banerjee2008} introduce the
Gaussian predictive process.
They project the original high-dimensional space onto a low-dimensional
subspace with a reduced set of locations,
fit a Gaussian process on this subspace and then use Kriging [\citet
{Krige1951}] to interpolate back to the original space.
\citet{Furrer2006} and \citet{Kaufman2008} reduce the computational
burden by covariance tapering.
Covariance tapering is a method whereby the covariance function is attenuated
with an appropriate compactly supported positive definite function such that
the covariance between pairs of sites that are farther apart than some
prespecified constant is set to zero.
This results in a sparse covariance matrix that can be inverted using
algorithms specifically designed for sparse matrices.
A different perspective is to view both the outcome and the coefficient
images as 3-dimensional functions,
and thus use the framework of functional data analysis [\citet{Ramsay2006}].
In particular, function-on-scalar regression models regress functional
outcomes on scalar predictors (covariates) [\citet{Reiss2010}].
\citet{Reiss2009} propose a functional principle component analysis
approach for
scalar outcome generalized linear models with functional predictors and
spatially varying coefficients.
\citet{Crainiceanu2009} introduce generalized multilevel functional
regression that
uses a truncated Karhunen--Lo\`{e}ve expansion to estimate spatially
varying coefficients.
All these alternative approaches rely on data reduction methods or
approximations to the processes.
In contrast, our model does not rely on data reduction methods or
approximations.
Parallel computing on a graphical processing unit (GPU) handles the
computational burden.

In the next two sections we formulate our Bayesian spatial model and
discuss some important algorithmic issues.
In Section~\ref{section-app} we apply our model to binary lesion maps
derived from $T_{2}$-weighted, high-resolution MRI images
from 250 subjects categorized into the five clinical subtypes of MS.
We compare our results with a mass univariate logistic regression
approach, Firth regression [\citet{Firth1993,Heinze2002}],
in terms of both parameter estimates and predictive performance.
Results from a simulation study are reported in Section~\ref{section-simstudy}.
We conclude the paper with a brief discussion.
Gibbs sampler details and some theoretical aspects of our model are
given in the supplemental article [\citet{Ge2014}].

\section{Spatial generalized linear mixed models}

Spatial generalized linear mixed models are similar to generalized
linear mixed models in that
both have a link function, fixed and random components.
The difference lies in how both the data and systematic component are
functions of space in the former.
We have binary data $Y(s)$ for each subject, where $s \in\mathbb
R^{d}$, for $d \geqslant1$ dimensional space (we work exclusively with $d=3$).
The link function is a monotonic function that relates the expectation
of the random outcome to the systematic component.
The systematic component relates a scalar $\eta(s)$ to a linear
combination of the covariates:
$\eta(s)={\mathbf x}^{\mathrm T}(s){\bolds\beta}(s)$.
That is, the covariates, parameters and $\eta$ are functions of space.
This representation of the systematic component is general enough to cover
spatially varying coefficients, spatially varying covariates, spatially
constant covariates and coefficients, and any combination of the above.
Typically for binary data, either the canonical link, the logit link,
with the natural parameter, the log odds or the probit link is used.
For computational reasons, we use the probit link (see Section~\ref{section-algo}).

This model, along with appropriate prior distributions for the model parameters,
applies to a wide range of scenarios with spatial binary data on a lattice,
though we focus only on our neuroimaging application.

\subsection{The model} \label{model}

We use notation from the spatial literature and refer to each voxel in
the image as a site. 
Let $s_{j}, j=1, \ldots, M$, denote the $j$th site within the brain
$\mathcal{B} \subset\mathbb{R}^{3}$,
where the sites are ordered lexicographically.
For subject $i=1, \ldots, N$ at site $s_{j}$,
let $Y_{i}(s_{j})$ denote a Bernoulli random variable representing
the presence [$Y_{i}(s_{j})=1$] or absence [$Y_{i}(s_{j})=0$] of a lesion.
For subject $i$, let ${\mathbf x}_{i}$ denote a column vector of $P$
subject-specific covariates
and let $w(s_{j})$ denote a single spatially varying covariate
evaluated at site $s_{j}$ that is shared among all subjects.
Our spatial generalized linear mixed model at site $s_{j}$ can then be
written as
%
\begin{eqnarray}
\bigl[Y_{i}(s_{j}) \mid p_{i}(s_{j})
\bigr] &\sim& \operatorname {Bernoulli}\bigl[p_{i}(s_{j})\bigr],
\label{sglmm:random}
\\
\Phi^{-1} \bigl\{\mathrm{E} \bigl[Y_{i}(s_{j})
\mid p_{i}(s_{j}) \bigr] \bigr\} &=& \eta_{i}(s_{j}),
\label{sglmm:link}
\\
\eta_{i}(s_{j}) &=& {\mathbf x}_{i}^{\mathrm T}
\bigl[\bolds\alpha+\bolds\beta (s_{j}) \bigr]+w(s_{j}){\gamma},
\label{sglmm:systematic}
\end{eqnarray}
reflecting the random, link and systematic component, respectively.

The random component is specified in \eqref{sglmm:random},
$Y_{i}(s_{j})$ is a Bernoulli random variable where $\Pr
[Y_{i}(s_{j})=1 ]=p_{i}(s_{j})$.
The link function is the probit link function, $\Phi^{-1}$, and the
systematic component is given by equation \eqref{sglmm:systematic}.
The motivation for this specific choice of systematic component will
become clear in Section~\ref{section-app}.
Since the expectation in \eqref{sglmm:link} is equal to the probability
that $Y_{i}(s_{j})=1$, $p_{i}(s_{j})$,
we can combine these three components into a \emph{spatial probit
regression model} with mixed effects:
%
\begin{equation}
\label{probit} \Phi^{-1} \bigl\{\Pr \bigl[Y_{i}(s_{j})=1
\mid\eta_i(s_{j}) \bigr] \bigr\}= {\mathbf
x}_{i}^{\mathrm T} \bigl[\bolds\alpha+\bolds\beta(s_{j})
\bigr]+w(s_{j}){\gamma}.
\end{equation}

The fixed effects in this model are the $P$-vector of parameters ${\bolds
\alpha}$ and the scalar parameter $\gamma$,
while the random effects are the $P$-vectors ${\bolds\beta}(s_{j})$, one
at each site.
Note that these random effects are spatially varying random effects and
not subject specific random effects.
Finally, $w(s_{j})$ is a covariate function of space, typically called
a spatially varying covariate,
while the spatially varying random effects are often called spatially
varying coefficients.
Note that our model is not implying a causal pathway.
Indeed, demyelination, that appears in $T_2$-weighted MRI imaging as
hyperintense lesions, may cause changes in both EDSS and PASAT.
Rather, our model is an association model, relating lesion prevalence
to covariates through the spatially varying coefficients.

We conclude our model specification by assigning priors to all parameters.
The fixed effect parameters have flat, improper, uninformative priors:
$\pi({\bolds\alpha}) \propto{\mathbf 1}$ and $\pi(\gamma) \propto1$, as is
standard for fixed effects regression parameters in Bayesian regression.
Spatial parameters have Markov random field or conditional
autoregressive model priors to account for the spatial structure.
Neighborhood systems of sites define these priors.
We regard two sites (i.e., voxels) as neighbors if they share a common
face and, thus, a site can have a maximum of six neighbors.
If sites $s_{j}$ and $s_{k}$ are neighbors, we write $s_{j} \sim s_{k}$,
and we denote the set of neighbors of site $s_{j}$ by $\partial
s_{j}= \{s_{k}\dvtx s_{k} \sim s_{j}  \}$
and the cardinality of this set by $n(s_{j})$.

The spatial random effect parameters have zero-centered multivariate
conditional autoregressive model (MCAR) priors as follows.
Define ${\bolds\beta}^{\mathrm T}= [{\bolds\beta}^{\mathrm T}(s_{1}),
\ldots,  {\bolds\beta}^{\mathrm T}(s_{M}) ]$: a $\mathrm{PM}$-length column vector.
Following the notation in \citet{Mardia1988}, the full conditional
distribution of ${\bolds\beta}(s_{j})$ is multivariate normal:
%
\begin{equation}
\bigl[{\bolds\beta}(s_{j}) \mid{\bolds\beta(-s_j)}, {\bolds
\Sigma} \bigr] \sim \operatorname{MVN} \biggl[\dispf{\sum_{s_{r}\in\partial s_{j}}{
\bolds\beta }(s_{r})} {n(s_{j})}, \frac{{\bolds\Sigma}}{n(s_{j})} \biggr],
\end{equation}
where ${\bolds\Sigma}$ is a $P\times P$ symmetric positive definite
matrix and $\bolds\beta(-s_{j})$ denotes the vector $\bolds\beta$
without the $P$ components at site $s_{j}$.
Note here that over most of the interior of the brain $n(s_{j})=6$ and
on the surface of the brain $n(s_{j})<6$.
Thus, this implies a spatially constant covariance over most of the brain.
In the discussion we show how this assumption can be relaxed.

By Brook's lemma [\citet{Brook1964}], the joint distribution, up to a
constant of proportionality, is
%
\begin{equation}
\pi [{\bolds\beta} \mid{\bolds\Sigma} ] \propto \exp \biggl\{-\dispf{1} {2}\sum
_{s_{i} \sim s_{j}} \bigl[{\bolds\beta}(s_{i})-{\bolds
\beta}(s_{j}) \bigr]^{\mathrm T}{{\bolds \Sigma}}^{-1} \bigl[{
\bolds\beta}(s_{i})-{\bolds\beta}(s_{j}) \bigr] \biggr\}.
\end{equation}
This joint distribution is improper and is not identifiable [\citet{Besag1986}],
as we can add an arbitrary constant to ${\bolds\beta}$ without changing
the joint distribution.
Nevertheless, as long as there is information in the data regarding
${\bolds\beta}$, the posterior of ${\bolds\beta}$ will be proper.
Last, we need to place a prior distribution on the hyperprior parameter
${\bolds\Sigma}$
or, equivalently, on the precision matrix ${\bolds\Sigma}^{-1}$.
We assign a Wishart prior with $\nu$ degrees of freedom and $P\times P$
identity scale matrix, ${\mathbf I}$, to the precision:
${\bolds\Sigma}^{-1} \sim W(\nu, {\mathbf I})$.
In the analysis below, we assign an improper,
uninformative prior to the precision matrix by setting $\nu=0$ while
noting that the posterior is proper.

\section{Some algorithmic issues} \label{section-algo}

The full conditional posterior distribution of ${\bolds\beta}(s_{j})$ is
not easy to sample.
We can resort to the Metropolis--Hastings algorithm [\citet
{Hastings1970}] or we can introduce continuous latent variables
that turn the spatial generalized linear mixed model into a spatial
linear mixed model [\citet{Albert1993}].
We adopt the latter approach, as then all full conditional posterior
distributions are easy to sample via Gibbs sampling [\citet
{Geman1984,Gelfand1990}].
We begin by introducing $N \times M$ independent continuous normal
latent variables $Z_{i}(s_{j}), i=1, \ldots, N$ and $j=1, \ldots, M$,
such that
%
\begin{equation}
\label{latent} \bigl[Z_{i}(s_{j}) \mid\eta_{i}(s_{j})
\bigr] \sim\mathrm{N} \bigl[\eta_{i}(s_{j}),1 \bigr].
\end{equation}
Now define the conditional probability that $Y_{i}(s_{j})=1$ given
$Z_{i}(s_{j})$ by
\[
\Pr \bigl[Y_{i}(s_{j})=1 \mid Z_{i}(s_{j})
\bigr]= \cases{ %
 1,&\quad $Z_{i}(s_{j})>0,$
\vspace*{2pt}\cr
0,&\quad $Z_{i}(s_{j})\leqslant0.$}
\]
The spatial linear mixed model is now given by \eqref{latent} and \eqref
{sglmm:systematic} along with the priors specified above.
All full conditional posterior distributions are now known
distributions that are easy to sample.
Thus, the joint posterior of all model parameters, given the latent
variables, are updated using a Gibbs sampler.

To show equivalence between the two models (the probit model and the
latent variable model),
we integrate out the latent variables to recover our probit model~\eqref
{probit}:
\begin{eqnarray*}& & \Pr \bigl[ Y_{i}(s_{j})=1
\mid\eta_i(s_{j}) \bigr]
\\
&&\qquad= \disp\int_{-\infty}^{+\infty}\Pr \bigl[Y_{i}(s_{j})=1
\mid z_{i}(s_{j}) \bigr]\pi \bigl[z_{i}(s_{j})
\mid\eta_i(s_{j}) \bigr]\,\mathrm{d}z_{i}(s_{j})
\\
&&\qquad= 1-\Phi \bigl[-\eta_i(s_{j}) \bigr]=\Phi \bigl[
\eta_{i}(s_{j}) \bigr].
\end{eqnarray*}

The full conditional distributions of the latent variables are
truncated normal distributions:
\[
\bigl[Z_{i}(s_{j}) \mid Y_{i}(s_{j}),
\eta_{i}(s_{j})\bigr] \sim \cases{ %
\mathrm{N}\bigl(\eta_{i}(s_{j}),1\bigr)
\times I\bigl(Z_{i}(s_{j})>0\bigr),&\quad $Y_{i}(s_{j})=1,$
\vspace*{2pt}\cr
\mathrm{N}\bigl(\eta_{i}(s_{j}),1\bigr)\times I
\bigl(Z_{i}(s_{j})<0\bigr),&\quad $Y_{i}(s_{j})=0,$}
\]
where $I(\cdot)$ is the indicator function. We use Robert's algorithm
[\citet{Robert1995}] to efficiently sample these full conditionals.
We provide all full conditional posterior distributions in the
supplemental article [\citet{Ge2014}].

\begin{figure}

\includegraphics{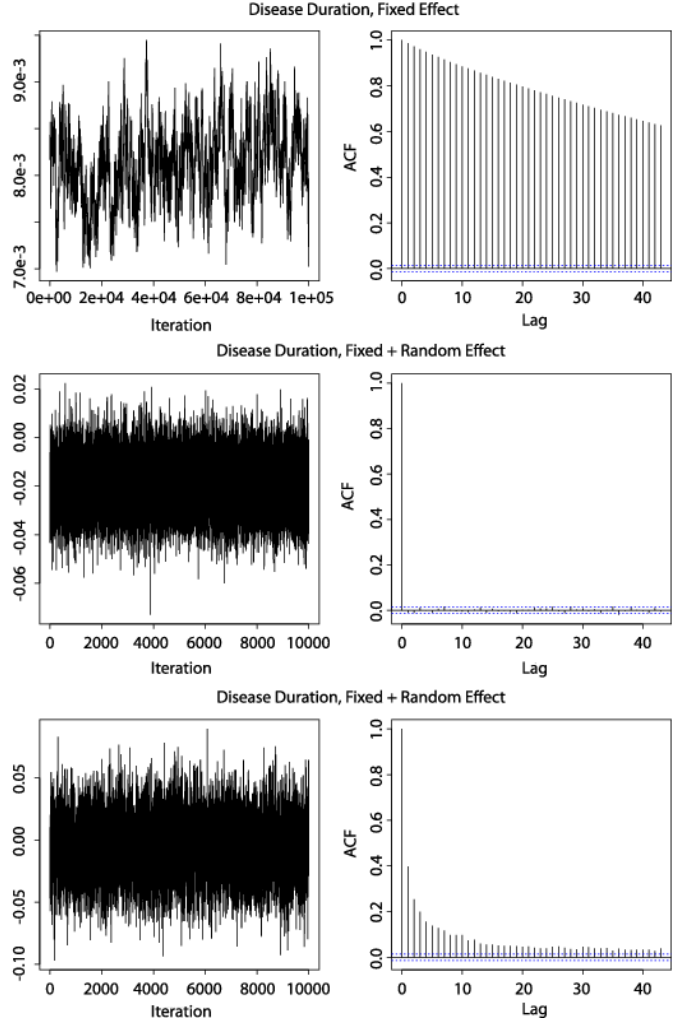}

\caption{Trace plots and autocorrelation functions (ACFs) for assessing
the mixing of our posterior sampler.
Top row: Disease duration fixed effect (spatially constant) after the
burn-in period, showing extremely strong autocorrelation.
Middle and bottom rows: After reparameterizing with sum of fixed and
random effects, the autocorrelation is reduced to acceptable levels;
middle row shows a voxel with large empirical probability of lesion,
while bottom row shows a voxel with an empirical probability of zero.
We found similar results for mixed effects estimates at randomly
selected locations in the brain.}
\label{fig-TS-ACF}
\end{figure}

Another issue is the extremely slow mixing (high autocorrelation) of
the fixed effects parameters ${\bolds\alpha}$,
as we initially observed and as reported by others [\citet{Gelfand2003}].
We accelerate the mixing by noting that our primary interest is not in
the fixed effects, ${\bolds\alpha}$,
but rather in the spatially varying coefficients, ${\bolds\alpha}+{\bolds
\beta}(s_{j})$.
We also note that the posterior variances of the components\vadjust{\goodbreak} of ${\bolds
\alpha}$ are much smaller
than the posterior variances of the components of the ${\bolds\beta}(s_{j})$
and that the ${\bolds\beta}(s_{j})$ do not suffer from slow mixing.
Thus, by reparameterizing the model with ${\bolds\beta}^{*}(s_{j})={\bolds
\alpha}+{\bolds\beta}(s_{j})$
and placing a nonzero-centered MCAR prior on the ${\bolds\beta}^{*}(s_{j})$,
we speed up mixing to acceptable levels (see time series and
autocorrelation function plots shown in Figure~\ref{fig-TS-ACF}).
One can easily recover the marginal posteriors of the components of
${\bolds\alpha}$
by simply taking the average of each component of ${\bolds\beta
}^{*}(s_{j})$ over $\mathcal{B}$
during the posterior simulation, that is, at each iteration of the
Gibbs sampling algorithm.

The final issue is the sheer size of both data and parameters.
In our application data set there are $N=250$ subjects,
with $M=274\mbox{,}596$ observed Bernoulli random variables per subject for a
total of 68,649,000 observations.
The length of each vector ${\bolds\beta}^*(s_{j})$ in our application is 10.
Thus, the total number of spatially varying coefficients that we need
to estimate is 2,745,960
along with the $10\times10$ covariance matrix ${\bolds\Sigma}$.
Therefore, simulating from the full posterior is an onerous task.
We reduce this computational burden by coding the problem to run in
parallel on a GPU.

We use a NVIDIA Tesla C2050 GPU card that has 3~Gb of main memory and
448 threads (independent processing units).
All data and code fit in 522 Mb of memory using floating point memory
for real-valued variables.
We run the algorithm for 150 thousand iterations, discarding the first
50 thousand as burn-in,
at which time the Markov chain has reached its stationary distribution.
The algorithm runs in just under 8 hours real time.
This is approximately an increase in speed of 45 times:
decreasing the real time from 15 days of computing on a single CPU (on
a 3~GHz Intel processor, coded in C$++$)
to just 8 hours of computing on the GPU.

One trick is necessary when updating the ${\bolds\beta}^{*}(s_{j})$, as
they are not independent,
creating a problem when parallelizing the code.
By leveraging the a priori and a posteriori conditional independence of
these vectors, we break the problem into two independent parts.
We explain this for the case of a 2-dimensional image, though the
extension to 3-dimensions is trivial (and is what we use in practice).
The pixels in a 2-dimensional image can be thought of as squares on a
checkerboard, alternately colored black and red.
Given the first order neighborhood system,
the parameters, ${\bolds\beta}^{*}(s_{j})$, on the red squares are
conditionally independent given the parameters on the black squares.
Similarly, the parameters on the black squares are conditionally
independent given the parameters on the red squares.
However, neighbors are dependent on one another.
Thus, to parallelize this problem, we divide and conquer, by extracting
all ``black square'' parameter vectors and updating them in parallel
conditional on the current parameter vector values on the red squares.
Likewise, we extract the ``red square'' parameter vectors and update
them in parallel
conditional on the current parameter values on the black squares.
This divide-and-conquer scheme respects the dependence of neighbors
during the parameter updates
and, thus, the entire algorithm remains a valid Gibbs updating algorithm.

\section{Analysis of MS lesions} \label{section-app}

Our motivating data set consists of 250 MS patients, each classified
into one of five clinical subtypes of MS.
In increasing order of clinical severity, these subtypes are clinically
isolated syndrome (CIS, 11 subjects),\vadjust{\goodbreak}
relapsed remitting (RLRM, 173 subjects), primary progressive (PRP, 13 subjects),
secondary chronic progressive (SCP, 43 subjects) and primary relapsing
(PRL, 10 subjects).
We note that CIS is not a true MS subtype, but rather is the first
clinical sign that MS may be imminent.
Between 30--70\% of patients diagnosed with CIS go on to develop MS
[\citet{Miller2005,Compston2008}].
Lesions were identified on $T_{2}$-weighted images in native
resolution, $0.977\times0.977\times3.000\ \mathrm{mm}^{3}$.
Two neuropathologists independently outline lesions on the MRI scans
using a semi-automated approach,
and a third radiologist mediates any discrepancies.
The result is a binary lesion map with 1 indicating the presence of a
lesion and 0 the absence of a lesion at each voxel.
The images are then affine registered to the Montreal Neurological
Institute (MNI) template
at $1\times1\times1\ \mathrm{mm}^{3}$ resolution using trilinear interpolation,
and thresholded at 0.5 to retain binary values.
To reduce the overall size of the images, over 2 million voxels,
we subsample every other voxel in each of the $x$-, $y$- and $z$-directions,
resulting in binary images with voxel size $2\times2\times2\ \mathrm{mm}^{3}$
for a total of $M=274\mbox{,}596$ voxels.

Finally, we note that our model is not dependent on the method of
lesion identification
and will work with any type of atlas-registered binary image data
exhibiting spatial dependence.

In the analysis we use six patient specific covariates:
clinical subtype (coded as five dummy variables), age, gender, DD, EDSS
score, PASAT score
and one spatially varying covariate shared by all subjects, the white
matter probability map.
The EDSS score is an ordinal measure of overall disability, ranging
from zero to ten in increments of one half [\citet{Kurtzke1983}].
The PASAT score is a neuropsychological test that assesses the capacity
and rate of information processing
as well as sustained and divided attention [\citet{Spreen1998}].
We treat clinical subtype as a nominal variable.
Subtype classification is based on the clinical course of the disease.
Patients classified as RLRM may convert to SCP, but, in general,
patients do not progress through the five disease subtypes
and, thus, we do not consider subtype as ordinal.
The white matter probability map, $w(s_{j})$, is the sole spatially
varying covariate.
MS is primarily a white matter disease, yet due to imperfect
inter-subject registration of the brain images,
each subject's white matter voxels will not perfectly overlap.
Thus, instead of constraining our analysis to a set of voxels defining
white matter,
we choose to analyze all brain voxels and use the white matter spatial covariate
to account for the gross differences in lesion incidence over the brain.

Thus, the covariate vector ${\mathbf x}_{i}$ has ten components. Associated
with each component is a spatially varying coefficient.
The first five are the intercepts for the five subtypes,
and the remaining are the slopes for age, gender, DD, EDSS score and
PASAT score.
We do not model interactions between subtypes and covariates,
as some subtypes have very little data (e.g., CIS with 11 subjects).
We mean-center age, DD, EDSS and PASAT scores prior to the analysis.

\subsection{Estimation}

We estimate the posterior distribution via Markov chain Monte Carlo (MCMC).
In particular, since all full conditional distributions have closed
form, we use the Gibbs sampler.
We simulate 100,000 draws from the posterior after a burn-in of 50,000,
by which time
the chain has converged to its stationary distribution, the posterior.

\begin{figure}

\includegraphics{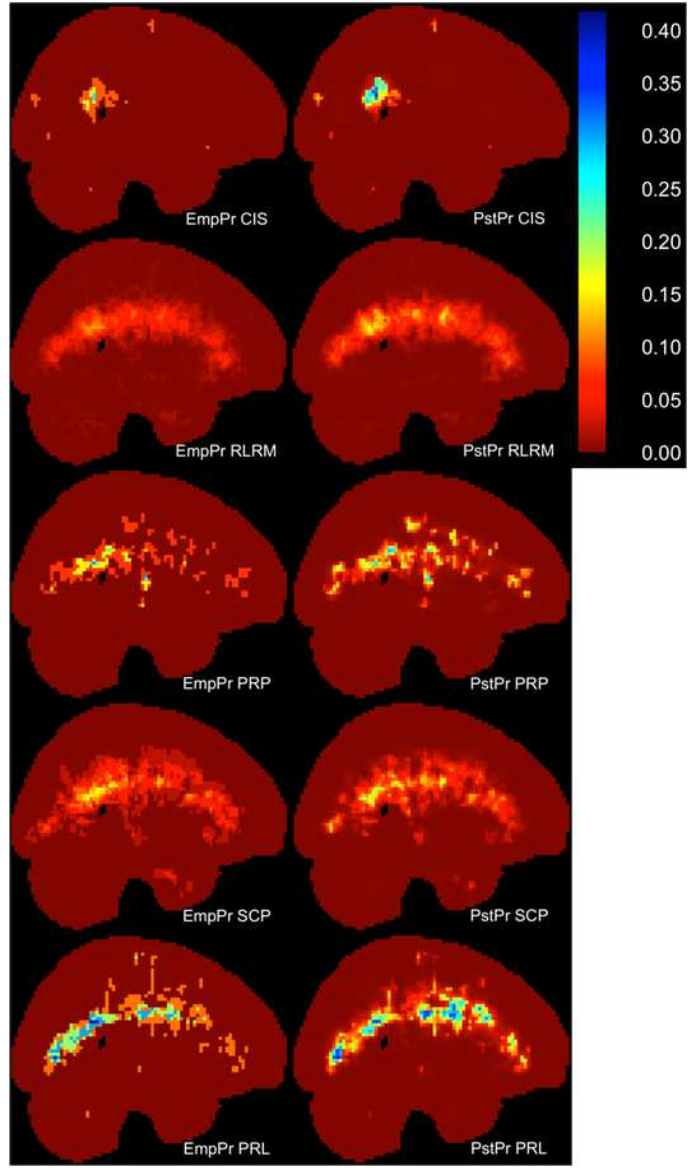}

\caption{Comparison of the empirical probabilities (left) and the
estimated mean posterior probabilities from our model (right)
for each of the five MS subtypes. Model estimates exhibit greater
smoothness due to our spatial MCAR prior.}\label{fig-Prbs}
\end{figure}

Figure~\ref{fig-Prbs} (left) shows the empirical lesion probabilities
for the five MS subtypes.
RLRM and SCP appear to have the most spatially extensive distribution
of lesions.
This, however, is an artifact of those groups having the most subjects.
Figure~\ref{fig-Prbs} (right) shows the estimated mean posterior
probabilities from our model.
Only the CIS patients show a dramatically different spatial
distribution of lesion incidence compared to the other subtypes.
This likely corresponds to the fact that CIS patients are those first
showing signs of having MS and thus have the lowest lesion load.
Furthermore, only 11 of the 250 subjects are classified as CIS.
However, other subtle differences are evident.
For example, PRL patients appear to have the highest overall lesion prevalence.

\begin{figure}[t!]

\includegraphics{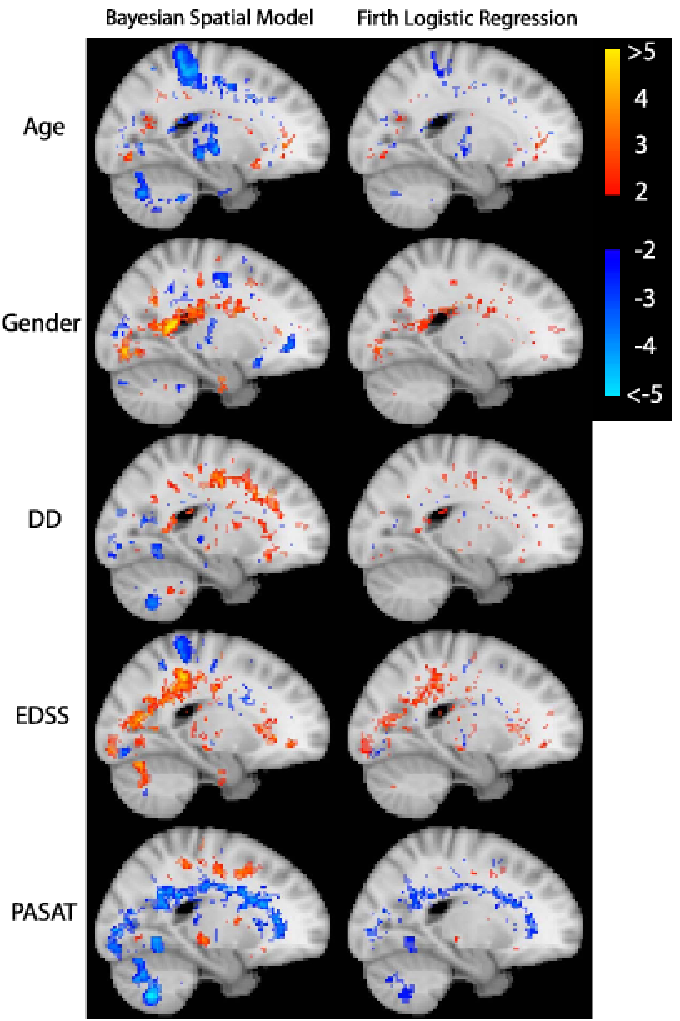}

\caption{A comparison of thresholded statistical significance maps for
covariates.
On the left are Bayesian standardized spatial maps (posterior mean
divided by posterior standard deviation)
for age, gender, disease duration, EDSS and PASAT scores,
and on the right are classical statistic spatial maps (mean divided by
standard deviation) from the mass univariate Firth logistic regression.
Color scale is set from 2 to 5 for positive values (values below 2 are
not shown, values of 5 or greater have maximal yellow color),
and from $-5$ to $-2$ for negative values (values above $-2$ are not shown,
values of $-5$ or smaller have the lightest blue color).
The statistic values from Firth logistic regression are significantly
attenuated and spatially contracted compared to our model estimates.}
\label{fig-SpatialCoef}\vspace*{-6pt}
\end{figure}

\begin{figure}

\includegraphics{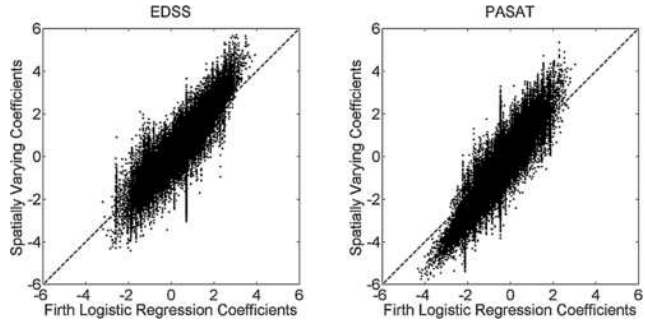}

\caption{Scatterplots of standardized parameter estimates from our
model versus those from Firth logistic regression on a voxel-by-voxel basis.
These plots show that statistic estimates from our model tend to have
greater magnitude at both large negative and positive values.
Vertical streaks reflect how many Firth statistic estimates are the
same at different voxels
(e.g., where there are no lesions), while our estimates vary.
Our model tends to spread out these coefficients due to the spatial
smoothing induced by the prior and, to a lesser extent, MCMC
error.}\label{fig-atten}
\end{figure}

Figure~\ref{fig-SpatialCoef} is a comparison of the thresholded (at
$\pm2$) statistical maps
(spatially varying coefficient estimates divided by their standard
deviations) for the covariates.
On the left are Bayesian standardized spatial maps (posterior mean
divided by posterior standard deviation)
for age, gender, DD, EDSS and PASAT scores, and on the right are
classical statistic spatial maps (mean divided by standard deviation)
from a mass univariate approach using Firth logistic regression [\citet
{Heinze2002,Firth1993}].
(Note that we compare with Firth regression as opposed to other
published methods [e.g., \citet{Kincses2011}]
that use a standard linear model to fit the binary data; as we state in
the \hyperref[sec1]{Introduction}, such linear models are inappropriate for binary data.)
Firth regression avoids the complete separation problem by shrinking
parameter estimates using a penalized likelihood approach.
For Firth regression, the mass univariate regressors include the five
dummy variables for clinical subtypes:
age, gender, DD, EDSS and PASAT scores.
Therefore, the Firth regression model shares the same subject specific
covariates with our Bayesian spatial model,
excluding the white matter spatial regressor, but of course ignores any
spatial dependence.
(While we could have included the white matter term, each voxel would
have a different estimate, whereas it is shared across all voxels in
our model.)
\begin{table}[t!]
\caption{Proportion of voxels that have standardized coefficients more
extreme than $\pm2$.
Our model results in a larger proportion of voxels that are
substantially large compared to Firth logistic regression.
This is due to the borrowing of strength from neighboring voxels in our
model}\label{tab-diff}
\begin{tabular*}{\textwidth}{@{\extracolsep{\fill}}lccccc@{}}
\hline
& \textbf{Gender} & \textbf{Age} & \textbf{DD} & \textbf{EDSS} & \textbf
{PASAT} \\
\hline
{Bayesian spatial model} & 3.78\% & 5.17\% & 4.73\% & 4.08\% &
5.08\% \\
{Firth logistic regression} & 0.75\% & 1.43\% & 1.11\% & 1.80\%
& 1.76\% \\
\hline
\end{tabular*}
\end{table}
Firth regression is performed in R with the \texttt{logistf} package (\url
{http://cran.r-project.org/web/packages/logistf}).
In Figure~\ref{fig-SpatialCoef}, one can see that the standardized
parameters from our model
are substantially larger and more spatially extensive compared to those
from Firth regression.
Table~\ref{tab-diff} numerically contrasts the extent of spatial
differences between our model and Firth regression.
The scatterplots of standardized parameter estimates in Figure~\ref{fig-atten} show the strengthening of these estimates.
For EDSS, for large positive coefficients, standardized parameter
estimates from our spatial model
tend to be larger than those from Firth regression.
Likewise, for PASAT, both for large positive and negative coefficients,
standardized parameter estimates from our model tend to be larger than
those from Firth regression.
This is a direct consequence of the MCAR prior, allowing parameter
estimates from neighboring voxels to borrow strength from one another,
producing smaller posterior estimates of the parameter variances.
The final result is an increase in standardized parameter estimates and
larger spatial extent of the signal.

\begin{figure}

\includegraphics{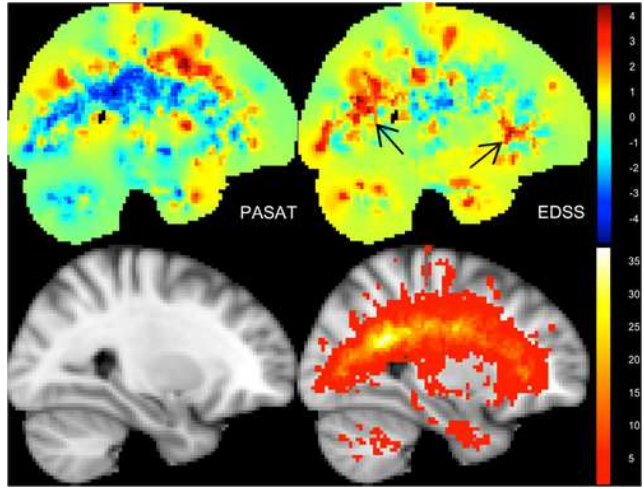}

\caption{A comparison of statistical significance maps for PASAT and
EDSS covariates.
A~single sagittal slice showing the standardized PASAT and EDSS
parameter estimates (top left and right, resp.).
High PASAT scores correspond to less damage from MS,
and hence negative correlation between PASAT score and lesion
occurrence can be seen, especially along the corpus callosum.
Higher EDSS scores correspond to more severe MS,
and hence the positive correlation between EDSS score and lesion
occurrence in the minor and major forceps
(anterior and posterior medial white matter tracks, see arrows in the
figure) can be seen.
For reference, the bottom panel shows the reference $T_{1}$ MRI
template (left)
and the same with empirical counts overlaid (right).}\label
{fig-PASAT-EDSS}
\end{figure}

We consider age, gender and disease duration as nuisance parameters.
Our main interest is in the EDSS and PASAT scores.
Figure~\ref{fig-PASAT-EDSS} shows a single sagittal slice of
standardized PASAT and EDSS parameter estimates (top left and right,
resp.).
For reference, the bottom panel shows the reference MRI template (left)
and empirical lesion counts overlaid on the reference template (right).
PASAT scores are negatively correlated (blue voxels, top left) with
lesion occurrence
as evident throughout areas of high lesion counts (lower PASAT scores
correspond to greater mental deficits).
EDSS scores are positively correlated (red voxels, top right) with
lesion occurrence within the minor and major forceps
(anterior and posterior medial white matter tracks that connect the
prefrontal cortex between the two hemispheres of the brain and
the parietal/occipital lobes between the two hemispheres, resp.),
which is consistent with higher EDSS scores corresponding to more
severe MS.
To the best of our knowledge, these findings have not been previously reported.

\subsection{Prediction}

Using our detailed Bayesian model, it is straightforward to make
predictions using Bayes' theorem. In particular, there is immense
potential clinical value in predicting an individual's MS subtype based
on their MRI lesion map, along with their age, gender, DD, EDSS and
PASAT scores.
To assess the predictive capabilities of our model,
we use a cross-validation approach, leaving one subject out at a time.
Direct implementation of leave-one-out cross-validation (LOOCV) would
be very time consuming,
as omitting one subject and rerunning the model would take 8 hours for
each of 250 subjects.
Thus, we adopt an importance sampling approach originally proposed by
\citet{Gelfand1992} where we need to run the model only once:
We remove each subject's contribution to the model by adjusting the
posterior at each iteration with an importance sample,
thus allowing held-out predictions for that subject.
We provide details of this importance sampling approach in the
supplemental article [\citet{Ge2014}].
We assign, a priori, a categorical distribution to clinical subtype
with equal probability of 0.2 to each of the 5 subtypes.

\begin{table}
\caption{The confusion matrices of the LOOCV classification using our
Bayesian spatial model (top) based on all in-mask voxels,
compared to a Na\"{i}ve Bayesian classifier (NBC) (middle) and Firth
logistic regression (bottom)
based on voxels that have at least two lesions across subjects.
Equal prior probability is assigned to each subtype when using the
Bayesian model.
The true subtype is shown in each row and the estimated subtype is
shown in each column.
The overall and the average classification rates for our Bayesian model
are $0.772 \pm0.052$ and $0.828 \pm0.047$, respectively.
The overall and the average classification rates for NBC are $0.552\pm
0.062$ and $0.245\pm0.053$, respectively.
The overall and the average classification rates for Firth logistic
regression are $0.672 \pm0.058$ and $0.300\pm0.057$, respectively}
\label{tab-LOOCV}
\begin{tabular*}{\textwidth}{@{\extracolsep{\fill}}lccccc@{}}
\hline
& \textbf{CIS} & \textbf{RLRM} & \textbf{PRP} & \textbf{SCP} &
\textbf{PRL} \\
\hline
\multicolumn{6}{c}{\textit{The Bayesian spatial model}} \\
{CIS} & \textbf{1.000} & 0.000 & 0.000 & 0.000 & 0.000 \\
{RLRM} & 0.243 & \textbf{0.734} & 0.000 & 0.023 & 0.000 \\
{PRP} & 0.154 & 0.000 & \textbf{0.846} & 0.000 & 0.000 \\
{SCP} & 0.140 & 0.000 & 0.000 & \textbf{0.860} & 0.000 \\
{PRL} & 0.100 & 0.000 & 0.100 & 0.100 & \textbf{0.700} \\[3pt]
\multicolumn{6}{c}{\textit{Na\"{i}ve Bayesian classifier}} \\
{CIS} & \textbf{0.000} & 1.000 & 0.000 & 0.000 & 0.000 \\
{RLRM}& 0.046 & \textbf{0.757} & 0.017 & 0.093 & 0.087 \\
{PRP} & 0.077 & 0.769 & \textbf{0.000} & 0.077 & 0.077 \\
{SCP} & 0.023 & 0.744 & 0.023 & \textbf{0.070} & 0.140 \\
{PRL} & 0.000 & 0.600 & 0.000 & 0.000 & \textbf{0.400} \\[3pt]
\multicolumn{6}{c}{\textit{Firth logistic regression}} \\
{CIS} & \textbf{0.000} & 1.000 & 0.000 & 0.000 & 0.000 \\
{RLRM} & 0.052 & \textbf{0.821} & 0.006 & 0.087 & 0.034 \\
{PRP} & 0.000 & 0.538 & \textbf{0.000} & 0.385 & 0.077 \\
{SCP} & 0.000 & 0.302 & 0.023 & \textbf{0.582} & 0.093 \\
{PRL} & 0.000 & 0.400 & 0.000 & 0.500 & \textbf{0.100} \\
\hline
\end{tabular*}
\end{table}

Table~\ref{tab-LOOCV} (top) shows the LOOCV classification results from
our model.
The rows show the true clinical subtype, while the columns show our
predicted subtype.
The overall correct classification rate is $0.772 \pm0.052$ ($X \pm
0.052$ denotes the limits of an approximate 95\% confidence interval
centered at $X$
based on a normal approximation to a binomial sample proportion).
The average classification rate, the unweighted average of the
per-subtype correct classification rates, is $0.828 \pm0.047$.
Due to the imbalance in group sizes, we find the average classification
rate is much more interpretable than the overall correct classification
rate. Consider a simple, obviously poor classifier that classifies
every one of the 250 subjects as RLRM (when in fact only 173 subjects
have this subtype). The overall correct classification rate in this
case is $173/250=0.692$, while the average classification rate is
$0.2$. The average classification rate balances out extremely high
correct classification rates in one or two subtypes that have the
largest samples sizes with extremely low correct classification rates
in subtypes that have very few subjects.
We see in Table~\ref{tab-LOOCV} (top) that if there is a
misclassification, that misclassification tends to be in the CIS subtype.
We investigated this further and found that those patients that are
misclassified to CIS tend to have fewer and smaller lesions
than those correctly classified (see Figure S1 
in the supplemental article [\citet{Ge2014}]).

As a comparison, we also perform LOOCV using a na\"{i}ve Bayesian
classifier (NBC) and Firth logistic regression.
Both NBC and Firth logistic regression assumes all voxels are mutually
independent, ignoring spatial dependence,
but NBC bases predictions on the empirical lesion rates alone (see
supplemental article [\citet{Ge2014}] for NBC details).
While assuming spatial independence seems like a gross oversimplification,
empirically NBC often outperforms more sophisticated and
computationally expensive approaches
[and there are theoretical arguments for this; see \citet{Zhang2004}].
Table~\ref{tab-LOOCV} (middle and bottom) shows the NBC and Firth
regression LOOCV classification results,
based on only those voxels that have at least two lesions across all subjects.
This ensures that for each voxel, after leaving one subject out,
there is at least one lesion in the remaining subjects (classification
based on all in-mask voxels produced much worse results).
The results of the NBC and Firth regression are largely biased to the
RLRM subtype.
The overall and the average correct classification rates for the NBC
are $0.552 \pm0.062$ and $0.245 \pm0.053$, respectively.
The overall and the average correct classification rates for Firth
regression are $0.672 \pm0.058$ and $0.300 \pm0.057$, respectively.
Despite the theoretical reasons offered by \citet{Zhang2004}, for this
data set,
our modeling approach significantly outperforms NBC in correctly
classifying subtype,
and it significantly outperforms Firth regression as well.

Although our model tends to misclassify a few patients into the
somewhat milder CIS subtype than the other methods,
it is much better at correctly classifying patients in the other four
subtypes, particularly PRL patients, than the other methods
(0.70 versus 0.40 and 0.10; cf. the last entry in each panel).
Furthermore, our overall correct and average classification rates are
much higher than
either the NBC approach or Firth logistic regression.
Both of these methods tend to classify most subjects into the RLRM, the
subtype with the largest number of patients.

Finally, to confirm that it is the imaging data and not just
demographic and clinical variables that are driving prediction,
we use a polytomous logistic regression (baseline categories model)
with no lesion data to perform this same classification.
We found accuracy rates of $0.776 \pm0.052$ (overall) and $0.419 \pm
0.061$ (average), demonstrating that it is the imaging data driving
prediction accuracy.

\subsection{Model diagnostics}

\begin{figure}

\includegraphics{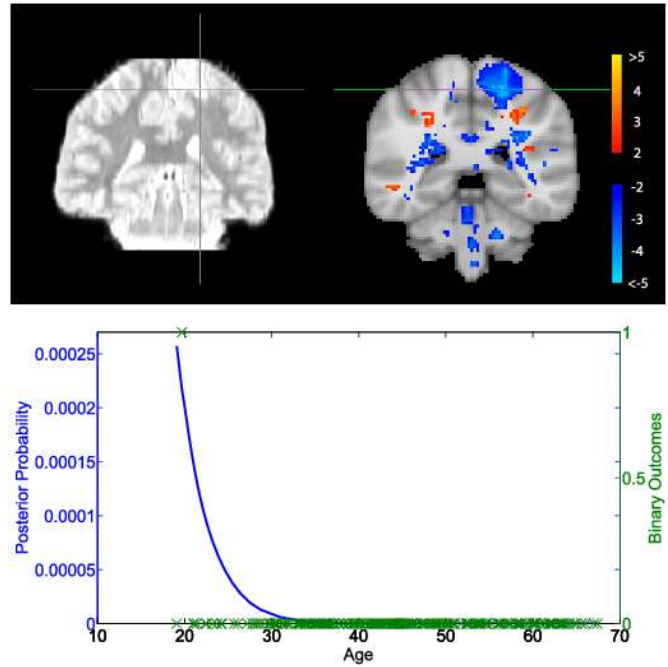}

\caption{Upper panel: the proton density image (left) of the outlying subject
and the standardized age coefficients overlaid on the brain atlas (right).
The lesion is seen on the proton density image and the standardized age
coefficients (above and below the thresholds, 2 and $-2$, are shown).
Lower panel: the fitted posterior lesion probabilities (blue line) as a
function of age for a selected voxel (see cross-hairs in upper panel)
in ``average'' RLRM females (gender, disease duration, EDSS and PASAT
scores all set to 0).
Also shown are the responses for all individuals at the selected voxel.}
\label{fig-Outlier}
\end{figure}

As with any regression analyses, model diagnostics should be performed.
For binary regression models these include investigation of outlying
and influential observations.
This should be done for each covariate at each voxel for each subject.
However, the sheer size of the problem and data make this untenable.
However, careful scrutiny of the coefficient maps, together with
estimated mean posterior probability maps, revealed a potential outlier.
Figure~\ref{fig-Outlier} shows a coronal view of a proton density image
(upper left)
and the standardized (posterior mean divided by posterior standard
deviation) coefficient map (upper right) for age (thresholded at $\pm2$).
The region in question is demarcated by cross-hairs and was identified
by its location near the superior cortical gray matter.
Although this region has large negative standardized coefficients
(voxel at cross-hairs, $-4.3$),
the posterior mean coefficient is only $-0.076$ and the mean posterior
probability is only $2.5 \times10^{-4}$ (bottom panel).
We find that there is one subject with a lesion in this location and
she is the second youngest patient in the data set
and has no discernible clinical disabilities from her disease.
An investigation of her images reveals that there is indeed a lesion
located in this region.
Thus, although there is strong statistical evidence that younger
patients are more likely to have a lesion in this location,
there is little scientific significance (an increase in probability of
about $2.5 \times10^{-4}$ over a subject one decade older).

Each area that may be of interest should be carefully examined, as we
have above.
This is true for all imaging based models.
We therefore caution that the results require careful interpretation
along with the empirical and/or posterior probability maps
to determine if the results are reliable or are simply the result of an
error in marking of a lesion.
The area of model diagnostics for large imaging problems is an open
problem that requires further work
(not only for our model, but for large imaging problems in general),
where traditional model diagnostics methods,
that rely heavily on graphical outputs, are not feasible.

\subsection{Convergence diagnostics}

MCMC algorithms must be monitored for convergence.
This is typically done by saving the chains for all parameters and
assessing convergence either visually or by Markov chain diagnostic methods.
Obviously, monitoring the approximately 2.75 million parameters in our
model is not feasible.
Thus, we selected 10 voxels where we monitor convergence.
Some of these voxels are located in regions of high lesion prevalence
and others in low lesion prevalence.
We ran the model from three random initial parameter settings.
Convergence was assessed using the Gelman--Rubin convergence diagnostic
for multiple chains [\citet{Gelman1992}].
The largest scale reduction factor observed was 1.01, indicating convergence.
As another check, we examined the 5 posterior mean coefficient maps of
interest (age, gender, disease duration, EDSS and PASAT scores)
and searched for the largest difference (in absolute value) between the
three possible pairs of runs for each of the 5 coefficient maps.
After locating the voxel at which the maximum difference occurs, using
the same initial settings and seeds, we reran the 3 simulations,
saving the draws of the coefficients at these voxels.
Gelman--Rubin convergence diagnostics revealed a largest scale reduction
factor of 1.01, indicating convergence at each of these voxels as well.

\section{Simulation study} \label{section-simstudy}

\begin{figure}

\includegraphics{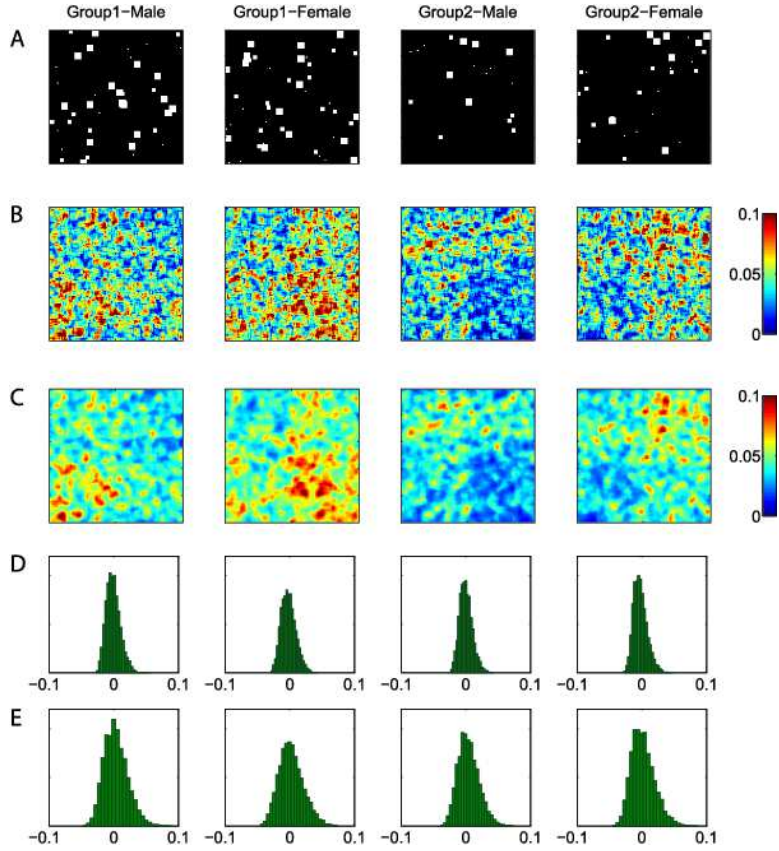}

\caption{Simulated data results.
\textup{(A)} Lesion patterns from randomly selected subjects.
\textup{(B)} Empirical lesion rates based on 100 randomly generated subjects.
\textup{(C)} Estimated lesion probabilities from the Bayesian spatial model.
\textup{(D)} Histograms of the difference between the estimated probabilities
from the Bayesian spatial model
and the true probability in the interior area of each quadrant.
\textup{(E)} Histograms of the difference between the estimated probabilities
from Firth logistic regression
and the true probability in the interior area of each quadrant.}
\label{fig-SimProb}
\end{figure}

We now present a simulation study to assess our model when ground truth
is known.
We create 2-dimensional, $100\times100$, images with different
behaviors in each of four $50\times50$ squares.
We assume that there are two groups of subjects consisting of both
males and females.
The number of lesions in each quadrant is drawn from a Poisson distribution.
On average, within the same group, females and males have the same
number of lesions on the left two quadrants,
while for each quadrant on the right, females have 4 more lesions than males.
Similarly, on average, for the same gender, subjects in groups 1 and 2
have the same number of lesions on the top two quadrants,
while for each quadrant at the bottom, subjects in group 1 have 4 more
lesions than subjects in group 2.
The locations of the lesions are uniformly distributed on each quadrant.
Each lesion is modeled as a square with side length a random variable
uniformly distributed on the set $\{1, 3, 5\}$.
Lesions are allowed to intersect with each other and merge into larger lesions.
Figure~\ref{fig-SimProb}(A) shows the binary images from some randomly
selected subjects.
For each combination of male vs. female and group 1 vs. group 2, we
simulated binary data for ten thousand subjects.
With the large number of subjects, an accurate estimation of true
lesion probability
can be obtained by calculating the empirical lesion rate at each pixel and
averaging over each quadrant excluding the outer two edge pixels on all
sides to reduce edge effects.
For example, the ``true'' lesion rates for the males in group 1 (the
first column in Figure~\ref{fig-SimProb}) are thus
0.0455, 0.0366, 0.0546 and 0.0459, clockwise, starting with the upper
left quadrant.

We then randomly selected 100 subjects from each combination, creating
a sample size of 400, and fitted our model.
The empirical lesion rates from the selected subjects are shown in
Figure~\ref{fig-SimProb}(B).
The regressors in the model are gender and two random intercepts
corresponding to the two groups.
All regressors are associated with spatially varying coefficients.
Females are coded 0 and males 1. We consider two pixels to be neighbors
if they shared a common edge.
The posterior distributions of the parameters are approximated by running
the Gibbs sampler for 12,000 iterations, discarding the first 2000 as burn-in.
Figure~\ref{fig-SimProb}(C) shows the estimated lesion probabilities from
our Bayesian spatial model.
Compared to the empirical rates in Figure~\ref{fig-SimProb}(B), the
smoothing effect is evident.
Figure~\ref{fig-SimProb}(D) shows histograms of the difference between
the estimated probabilities from the Bayesian spatial model
and the ``true'' lesion rates in each quadrant.
The mean squared error (MSE), averaged over all pixels, is $1.20\times10^{-4}$.
As a comparison, we also performed Firth logistic regression at each pixel.
The histograms of the difference between the estimated probabilities
from Firth logistic regression
and the ``true'' lesion rates in each quadrant are shown in Figure~\ref{fig-SimProb}(E).
These histograms are wider, and the MSE is $3.33\times10^{-4}$:
approximately 3 times larger.

\begin{figure}

\includegraphics{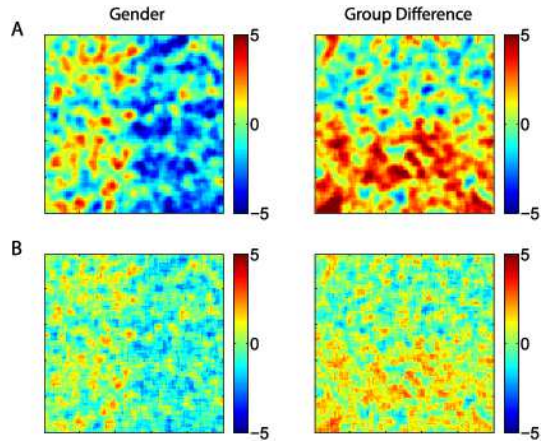}

\caption{Simulated data results. \textup{(A)} Standardized spatially varying
coefficients of gender and the difference between the two intercepts.
\textup{(B)} Standardized coefficient maps produced by Firth logistic regression.}
\label{fig-SimCoef}
\end{figure}

Figure~\ref{fig-SimCoef}(A) shows the standardized spatially varying
coefficients of gender
and the difference between the two intercepts (Group1--Group2).
The spatial coefficient maps clearly reflect the spatially varying
effect of gender and the group difference.
Figure~\ref{fig-SimCoef}(B) shows the standardized coefficient maps from
Firth logistic regression.
Without spatial regularization, the significance map is attenuated
relative to that from our model.
We note here that the true coefficient maps are not available and that
the comparisons made here are relative between the two models.

\section{Discussion} \label{Discussion}

In this paper we present a Bayesian spatial model that respects the
binary nature of the data and
exploits the spatial structure of MS lesion maps without use of an
arbitrary smoothing parameter.
The method is suitable to model any patterns of lesions, including
$T_{2}$ lesions, which show a variety of sizes and shapes, $T_{1}$
``black-hole'' lesions
and any other types of lesions from which a binary image marking the
location of the lesions can be derived.
By explicitly including covariates and allowing for spatially varying
coefficients,
our model provides spatial information that most current empirical
approaches cannot;
for example, we obtain estimates and estimator precisions for the
spatially varying effects of age, gender, DD, EDSS and PASAT.

Our model provides excellent classification accuracy for predicting
clinical subtypes of MS
based on the entire pattern of lesions over the brain, as well as
demographic and behavioral variables.
As noted in the \hyperref[sec1]{Introduction}, associations between MRI findings and
clinical outcomes have been paradoxically weak.
Hence, our construction of a model that not only finds disease subtype
differences but also provides high prediction accuracies
is an important advance for this area.
Specifically, we know of no other work that performs such 5-way
classification over disease subtypes.
It appears that, by borrowing strength from neighboring voxels and
respecting the binary nature of the data, our modeling approach
overcomes this paradox to some degree.

Our model is easily extended to include EDSS sub-scores (i.e.,
disabilities in 7 functional systems) or other diagnostic measures.
Last, in Section~\ref{model} we note that the covariance matrix was
constant over much of the brain.
This can easily be relaxed by allowing a voxel specific covariance matrix,
${\bolds\Sigma}_{j}$, at the expense of larger computational burden. %

The large data set analyzed in this paper, 3-dimensional images each
with about 275K voxels from 250 subjects,
consists of approximately 70 million observed outcomes, represents a
challenge for any spatial data analysis.
There are several spatial models, as reviewed in the \hyperref[sec1]{Introduction}, with
spatially varying coefficient processes which
in principle could be used, but, to our knowledge, have not been
applied to such a large problem.
Compared to these methods, our model does not require any approximation
or data reduction method.

One limitation of our data is the use of affine registration to align
subjects to a common space
and, thus, a future direction is to use high-dimensional nonlinear
registration that can better align brain structures across subjects.
We could then investigate whether the predictive accuracy or covariate maps
will be improved by the better structural alignment afforded by
nonlinear registration.
However, an issue with nonlinear registration is that lesion volumes
may not change proportionally as they do with affine registration.
For a given subject, some lesions may shrink by nonlinear registration
while others become larger.
When a lesion is shrunk, we are implicitly stating that this lesion is
less important for that subject.
Thus, an intriguing methodological direction is the development of a
model for binary lesion data
that accounts for local volume change induced by nonlinear registration,
as Voxel Based Morphometry [\citet{Ashburner2000}] does with its
Jacobian-based adjustment.

All code, both CPU and GPU versions, is available by contacting the
authors or online at \url{http://go.warwick.ac.uk/tenichols/BSGLMM}.

\section*{Acknowledgments}
We would also like to thank
Ernst-Wilhelm Rad\"{u} and the staff of the Medical Image Analysis
Center (MIAC)
at the University Hospital Basel, Switzerland for generously providing
the data and motivating this work. We also thank the
two anonymous reviewers, the Associate Editor and Editor for the many
constructive comments that have made this a much better manuscript.

\begin{supplement}[id=suppA]
\stitle{Supplement to
``Analysis of multiple sclerosis lesions via spatially varying coefficients''}
\slink[doi]{10.1214/14-AOAS718SUPP}  
\sdatatype{.pdf}
\sfilename{aoas718\_supp.pdf}
\sdescription{This supplement contains full details of the Gibbs
sampler, leave-one-out cross-validation and the
na\"{i}ve Bayesian classifier. It also contains supplementary
figures.}
\end{supplement}



\printaddresses

\end{document}